\documentstyle[multicol,aps,epsf]{revtex}
\begin{document}
\draft

\title{Diffusion-Limited Coalescence, $A+A\rightleftharpoons A$, with a Trap}
\author{Daniel ben-Avraham\footnote
  {{\bf e-mail:} qd00@polaris.clarkson.edu}}

\address{Physics Department, and Clarkson Institute for Statistical
Physics (CISP), \\ Clarkson University, Potsdam, NY 13699-5820}
\maketitle

\begin{abstract}
We study diffusion-limited coalescence, $A+A\rightleftharpoons A$, in
one dimension, and derive an exact solution for the steady state in the
presence of a trap.  Without the trap, the system arrives at an {\it
equilibrium\/} state which satisfies detailed balance, and can therefore be
analyzed by classical equilibrium methods.  The trap introduces an irreversible
element, and the stationary state is no longer an equilibrium state.  The
exact solution is compared to that of a reaction-diffusion equation
--- the habitual approximation method of choice.  The exact solution allows
us to determine the rate coefficients in the reaction-diffusion equation,
without appealing to renormalization group techniques.
\end{abstract}

\begin{multicols}{2}

\section{Introduction }
\label{sec:intro}
Non-equilibrium kinetics of diffusion-limited reactions has been the subject
of much recent interest
\cite{vanKampen,Haken,Nicolis,Ligget,reaction-reviews,JStatPhys}.  In contrast
to equilibrium systems --- which are best analyzed with standard thermodynamics
--- or reaction-limited processes --- whose kinetics is well described by
classical rate equations \cite{Laidler,Benson}--- there is no general approach
to non-equilibrium, diffusion-limited reactions.

In this letter we study a diffusion-limited coalescence process in one
dimension: $A+A\rightleftharpoons A$, which can be analyzed {\it exactly\/}
\cite{Coalescence,Doering}.  When a trap is introduced, the resulting steady
state is a non-equilibrium state.  We derive an exact description of this state
and compare it to the prediction from a reaction-diffusion equation --- the
standard approximation method of choice.  The exact solution allows us
to determine the appropriate rate constants of the reaction-diffusion equation
directly.  Normally, this feat requires a renormalization group analysis.

The rest of this letter is organized as follows. In
Section~\ref{sec:formulation} we present a lattice model of diffusion-limited
reversible coalescence, along with the exact method of analysis; the method of
Empty Intervals, also known as the method of Inter-Particle Distribution
Functions. The stationary state in the presence of a trap is derived and
analyzed in Section~\ref{sec:trap}.
In Section~\ref{sec:mean-field} we
compare the exact solution to that of a reaction-diffusion equation, and
devise strategies to determine the appropriate rate coefficients.  We conclude
with a discussion and open questions in Section~\ref{sec:discussion}.

\section{Reversible Coalescence}
\label{sec:formulation}

Our model \cite{Coalescence,Doering} is defined on a one-dimensional lattice of
lattice spacing
$a$.  Each site is in one of two states: occupied by a particle $A$, or empty.
Particles hop randomly into nearest neighbor sites, at rate
$D/a^2$.  A particle may give birth to an additional particle, into a
nearest neighbor site, at rate $v/2a$ (on either side of the particle).
If hopping or birth occurs into a site which is already occupied, the target
site
remains occupied.  The last rule means that coalescence, $A+A\to A$, takes
place
{\it immediately\/} upon encounter of any two particles.  Thus, together with
hopping and birth, the system models the diffusion-limited reaction process
$A+A\rightleftharpoons A$.

An exact treatment of the problem is possible through the method of
Inter-Particle Distribution Functions (IPDF).  The key concept is
$E_{n,m}(t)$ --- the probability that sites $n,n+1,\cdots,m$ are empty at time
$t$.  The probability that site $n$ is occupied is
\begin{equation}
\label{disc-conc}
{\rm Prob}({\rm site\ }n{\rm\ is\ occupied})=1-E_{n,n}\;.
\end{equation}
The event that sites $n$ through $m$ are empty (prob. $E_{n,m}$) consists of
two
cases: site $m+1$ is also empty (prob. $E_{n,m+1}$), or it is occupied.  Thus
the probability that sites $n$ through $m$ are empty, but site
$m+1$ is occupied is $E_{n,m}-E_{n,m+1}$.  With this (and with a similar rule
for
when the particle is to the left of the empty segment) one can write down
a rate equation for the evolution of the empty interval probabilities:
\begin{eqnarray}
\label{mastereq}
&&{\partial E_{n,m}\over\partial t}
   = {D\over a^2}(E_{n,m-1}-E_{n,m})\nonumber\\
   &&- {D\over a^2}(E_{n,m}-E_{n,m+1})\nonumber\\
   &&- {D\over a^2}(E_{n,m}-E_{n-1,m})\nonumber\\
   &&+ {D\over a^2}(E_{n+1,m}-E_{n,m})\nonumber\\
   &&- {v\over2a}[(E_{n,m}-E_{n,m+1})+(E_{n,m}-E_{n-1,m})]\;.
\end{eqnarray}
Eq.~(\ref{mastereq}) is valid for $m>n$.  The special case of $m=n$ yields the
boundary condition
\begin{equation}
\label{discBC}
E_{n,n-1}=1\;.
\end{equation}
The fact that the $\{E_{n,m}\}$ represent {\it probabilities\/} implies the
additional condition that $E_{n,m}\geq 0$.  Finally, if the system is not empty
then $E_{n,m}\to 0$ as $n\to-\infty$ and $m\to\infty$.

In many applications, it is simpler to pass to the continuum limit.  We write
$x=na$ and $y=ma$, and replace $E_{n,m}(t)$ with $E(x,y;t)$.  Letting $a\to 0$,
Eq.~(\ref{mastereq}) becomes
\begin{equation}
\label{eqE}
{\partial E\over\partial t}=D({\partial\over\partial x^2}
  +{\partial\over\partial y^2})E - {v\over 2}({\partial E\over\partial x}
  -{\partial E\over\partial y}) \;,
\end{equation}
with the boundary conditions,
\begin{eqnarray}
\label{BC1}
E(x,x;t) &=& 1\;,\\
\label{BCpositive}
E(x,y;t) &\geq& 0\;,\\
\label{BC0}
\lim_{{x\to-\infty\atop y\to+\infty}}E(x,y;t) &=& 0\;.
\end{eqnarray}

The concentration of particles is obtained using Eqs.~(\ref{disc-conc}) and
(\ref{discBC}), and passing to the continuum limit:
\begin{equation}
\label{conc}
c(x,t)=-{\partial E(x,y;t)\over\partial y}|_{y=x}\;.
\end{equation}
It can also be shown \cite{Coalescence,Doering} that the conditional joint
probability for having particles at $x$ and $y$ but none in between, is
\begin{equation}
\label{IPDF}
P(x,y;t)=-{\partial^2E(x,y;t)\over\partial x\,\partial y}\;.
\end{equation}
Given a particle at $x$, the probability that the next nearest particle is at
$y$, {\it i.e.}, the IPDF, is $p(x,y;t)=(1/c)P$.

\section{Steady state with a trap}
\label{sec:trap}
The steady state of Eq.~(\ref{eqE}), with the boundary conditions (\ref{BC1})
--
(\ref{BC0}), when $\partial E/\partial t=0$, is
\begin{equation}
\label{E_eq}
E_{eq}=e^{-{v\over2D}(y-x)}\;.
\end{equation}
This corresponds to the equilibrium concentration of particles (using
Eq.~\ref{conc})
\begin{equation}
\label{ceq}
c_{eq}={v\over2D}\equiv\gamma\;.
\end{equation}
Another trivial solution is $E(x,y)=1$; it represents a totally empty system.
But the solution~(\ref{E_eq}) is stable, while the vacuum state is not.  In
fact, when the initial state of the system is a mixture of the two phases:
$c(x,t\!=\!0)=0$ for $x<0$, and $c(x,t\!=\!0)=c_{eq}$ for $x>0$, say, then the
stable phase invades the unstable phase.  The front between the two phases
propagates at a constant speed, similar to the case of Fisher waves
\cite{waves}. Here we wish to study another inhomogeneous situation, where
there
is a perfectly absorbing trap at the origin instead of the initial empty
half-space.  The trap depletes its immediate neighborhood, but a non-trivial
steady state is expected as the depletion zone created by the trap is
continually replenished by a stream of particles from the stable phase.

To derive the appropriate boundary condition, we turn back to the discrete
representation.  In the presence of a perfect trap at $n=0$,
Eq.~(\ref{mastereq}) is then limited to $0<n<m$.  The special equation for
$n=0$ is
\begin{eqnarray}
{\partial E_{0,m}\over\partial t} &=&
{D\over a^2}(E_{0,m-1}+E_{0,m+1}+E_{1,m}-3E_{0,m})\nonumber\\
&-& {v\over 2a}(E_{0,m}-E_{0,m+1})\;.
\end{eqnarray}
Comparison to Eq.~(\ref{mastereq}) yields the discrete boundary condition
\begin{equation}
E_{-1,m}=E_{0,m}\;,
\end{equation}
which in the continuum limit becomes
\begin{equation}
\label{BCtrap}
{\partial E(x,y;t)\over\partial x}|_{x=0}=0\;.
\end{equation}
In addition, the boundary condition~(\ref{BC0}) is now replaced by
\begin{equation}
\label{BC0trap}
\lim_{y\to\infty}E(0,y;t) = 0\;.
\end{equation}

The stationary solution to Eq.~(\ref{eqE}), confined to the wedge ($0\leq
x\leq y$), which satisfies the boundary conditions (\ref{BC1}),
(\ref{BCpositive}), (\ref{BCtrap}) and (\ref{BC0trap}), is
\begin{equation}
\label{Es}
E_s(x,y)=e^{-\gamma(y-x)}+\gamma(y-x)e^{-\gamma y}\;.
\end{equation}
Far away from the trap, as $x,y\to\infty$, this converges to the equilibrium
result of Eq.~(\ref{E_eq}). From~(\ref{conc}), we obtain the stationary
concentration profile:
\begin{equation}
\label{cs-true}
c_s(x)=\gamma(1-e^{-\gamma x})\;.
\end{equation}
As expected, there is a depletion zone of size $1/\gamma=2D/v$ near the trap,
and the concentration grows asymptotically to $c_{eq}$ as $x\to\infty$.

The IPDF between nearest particles is surprising.  From Eqs.~(\ref{IPDF}) and
(\ref{Es}) we obtain the conditional joint probability
\begin{equation}
\label{jointPs}
P_s(x,y)=\gamma^2 e^{-\gamma y}(e^{\gamma x}-1)\;.
\end{equation}
Dividing $P_s(x,y)$ by $c_s(x)$ yields the ``forward" IPDF --- the
probability that given a particle at $x$ the next nearest particle {\it to its
right\/} is at $y$:
\begin{equation}
\label{ps}
p_s(z)=\gamma e^{-\gamma z}\;;\qquad z\equiv y-x\;.
\end{equation}
The notation chosen here emphasizes the unexpected result that $p_s$ is
translationally invariant.  What is more, this Poissonian IPDF is
characteristic of particles at {\it equilibrium\/}, when there are no
correlations between their various positions.  Indeed, exactly the same IPDF is
obtained for the equilibrium state of Eq.~(\ref{E_eq}), without the trap!

To obtain the ``backward" IPDF --- the probability that given a particle at
$y$ the next nearest particle {\it to its left\/} is at a distance $z$ --- we
divide $P_s(x,y)$ by $c_s(y)$:
\begin{equation}
q_y(z)=\gamma{e^{-\gamma z}-e^{-\gamma y}\over 1-e^{-\gamma y}}\;.
\end{equation}
The fact that $q_y(z)$ is not translationally invariant comes as no surprise,
because of the trap at $x=0$.  However, $q_y(z)$ does not normalize properly!
The reason for that is that there is a finite
chance that there are no particles between the particle at $y$ and
the trap; {\it i.e.}, the particle at $y$ is the nearest
particle to the trap.  The probability that this happens is $p_0(y)=-\partial
E/\partial y|_{x=0}$, or
\begin{equation}
p_0(y)=\gamma^2ye^{-\gamma y}
\end{equation}
With this understanding, the proper normalization condition is
\begin{equation}
c_s(y)\int_0^y q_y(z)dz+p_0(y)=c_s(y)\;,
\end{equation}
which is indeed met.

\section{Reaction-diffusion equation}
\label{sec:mean-field}
So far, we have presented an exact solution to the problem of
diffusion-limited reversible coalescence with a trap.  Exactly solvable
models of diffusion-limited reactions, however, are rare.  We now wish to
discuss one of the most widely used approximation methods, in light of the
exact
results.

The method is that of reaction-diffusion equations.  Here one assumes
the existence of a {\it mesoscopic\/} length scale within which the system is
homogeneous and well mixed, and where the reaction rates can be accounted for
as in classical rate equations.  At longer length scales, variations in the
concentration, $c(x,t)$, give rise to diffusion.  In our case,  the
appropriate reaction-diffusion equation is
\begin{equation}
\label{diff-reac}
{\partial c(x,t)\over\partial t}=D{\partial^2c\over\partial x^2}+k_1c-k_2c^2\;,
\end{equation}
where $k_1$ and $k_2$ represent the ``effective" rates of birth and
coalescence, respectively.  The trap at the origin imposes the boundary
condition:
\begin{equation}
\label{c=0}
c(0,t)=0\;.
\end{equation}

An alternative approach is that of writing down an infinite hierarchy of rate
equations for the $n$-point density correlation functions, and truncating the
hierarchy with a Kirkwood ansatz at some convenient stage.  In the simplest
case, one truncates the hierarchy at the level of single-point density
functions.  This is achieved by neglecting all correlations, and by expressing
multiple-point density functions as products of single-point densities.  In
view of the peculiar IPDF in our problem (Eq.~\ref{ps}), this seems a
promising approximation.

Let the probability of having a particle at site
$n$ be
$\rho_n(t)$, then the joint probability of having particles at both $n$ and
$m$ at time $t$ may be approximated as
$\rho^{(2)}_{n,m}(t)\approx\rho_n(t)\rho_m(t)$. In this fashion, our model is
described by the equation
\begin{eqnarray}
\label{rho}
{\partial\rho_n\over\partial t} = &&
 {D\over a^2}[-2\rho_n+(1-\rho_n)(\rho_{n-1}+\rho_{n+1})]\nonumber\\
&& +{v\over 2a}(1-\rho_n)(\rho_{n-1}+\rho_{n+1})\;,
\end{eqnarray}
and the boundary condition $\rho_0(t)=0$.
Notice that in the stationary limit, and without the trap, Eq.~(\ref{rho}) is
{\it exact\/}, since in the equilibrium state of the infinite system the
particles really are uncorrelated!  Indeed, solving~(\ref{rho}) when
$\partial\rho_n/\partial t=0$ yields $c=\rho/a=v/(2D+av)$, which agrees with
$c_{eq}$ of Eq.~(\ref{ceq}) when $a\to 0$.

It is tempting to try and connect Eq.~(\ref{rho}) to the reaction-diffusion
approach, by passing to the continuum limit.  This, however, does not work:
writing $c(x,t)=\rho_n(t)/a$, and letting $a\to 0$ while keeping only up to
first-order terms in $a$, yields
\begin{equation}
\label{rho_limit}
{\partial c\over\partial t}=D{\partial^2c\over\partial x^2}
  +{v\over a}c - ({2D\over a}+v)c^2\;.
\end{equation}
Thus, it is impossible to identify $k_1$ and $k_2$ in this manner, and one is
forced to work with the discrete equation~(\ref{rho}).  (An approximate
solution could be found by proceeding with Eq.~(\ref{rho_limit}) anyway,
ignoring the fact that $a$ is supposed to be infinitesimally small.)

We now use the exact solution of Section~\ref{sec:trap} to attempt and
determine $k_1$ and $k_2$.  We note that without the birth and coalescence
reactions the particles would simply diffuse with a diffusion constant
$D$ --- the same $D$ as in the hopping rate $D/a^2$ of the microscopic rules.

First consider the stationary solution of Eq.~(\ref{diff-reac}), for an
infinite system without the trap: $c_{eq}=k_1/k_2$.  To conform with the
exact solution of~(\ref{ceq}), we must have
\begin{equation}
\label{k1/k2}
{k_1\over k_2}={v\over2D}\;.
\end{equation}

The stationary solution to
Eq.~(\ref{diff-reac}) with the trap --- the boundary condition~(\ref{c=0})
--- is
\begin{equation}
\label{tanh}
{c_s(x)\over c_{\infty}}={3\over2}\tanh^2\Big(\sqrt{{k_1\over3D}}x
     +  \tanh^{-1}\sqrt{{1\over3}}\,\Big)-{1\over2}\;,
\end{equation}
where $c_{\infty}=c_{eq}$ is the concentration of particles infinitely far away
from the trap.

The concentration profile described by Eq.~(\ref{tanh}) looks
similar to the exact result of Eq.~(\ref{cs-true}).  One could now use
different criteria to further constrain $k_1$ and $k_2$. Demanding the same
asymptotic behavior far away from the trap;
$\lim_{x\to\infty}\ln[1-c_s(x)/c_{eq}]/x=-v/2D$, we get
\begin{equation}
k_1={3\over8}{v^2\over2D}\;.
\end{equation}
On the other hand, if we require the same behavior close to the trap;
$(\partial c_s/\partial x)_{x=0}=(v/2D)^2$, we get
\begin{equation}
k_1={9\over8}{v^2\over2D}\;.
\end{equation}
Clearly, it is impossible to fix the short range behavior and the long range
behavior simultaneously.  Instead, one may write
\begin{equation}
k_1=\alpha{v^2\over 2D}\;,{\rm\ \ and\ \ } k_2=\alpha v\;,
\end{equation}
where $\alpha$ is a single fitting parameter of order unity.  A
least square fit in the range $0\leq c_{eq}x\leq 5$ is achieved with
$\alpha=0.889\,$.

\section{Discussion}
\label{sec:discussion}
We have solved the problem of diffusion-limited reversible coalescence with a
trap, in one dimension, exactly.  The result is tantalizingly simple: the
stationary concentration profile is exponential.  Moreover, the distribution of
distances between nearest particles (the IPDF) is also exponential, similar to
that of particles in an {\it equilibrium\/} process --- the same process in the
absence of the trap.  This does not mean, however, that the
distribution of particles in the two cases is identical: it
is just a peculiarity of the IPDF in this particular model.  To be sure, the
distribution of particles in the equilibrium situation is fully random and
uncorrelated, whereas in the presence of the trap it is not!

We have also contrasted the exact solution with the alternative, traditional
approach of reaction-diffusion equations, highlighting the fact that the latter
is merely an approximation method.  Our model provides a clear example where
the
effective rates of the reaction-diffusion equation can be related to the
microscopic rates of the underlying process, without appealing to
renormalization.

An interesting open question is at what dimension reaction-diffusion
equations accurately describe the kinetics of the system.  Previously, we
had conducted numerical studies of Fisher waves in the coalescence
process, suggesting that the critical dimension is $d_c=3$ \cite{waves}.  The
present model could present an advantage in future numerical studies, because
of its non-trivial {\it stationary\/} state.

Future work will also include the investigation of two-point density
correlation functions, as well as the influence of a drift away from the
trap.  Both problems can be formulated rigorously within the framework of
the IPDF method \cite{Doering}.  Multiple-point density correlation functions
would shed further light on the breakdown of reaction-diffusion equations in
low
dimensions.  Drift away from the trap could potentially give rise to an
interesting phase transition: from a system with a non-trivial steady state
(such as in our case, when the drift is zero), to a system with only a trivial
steady state (the vacuum), as the drift increases beyond a critical point.

\acknowledgments
I thank L.~Glasser, P.~Krapivsky, S.~Redner, and L.~Schulman for valuable
discussions.

\end{multicols}


\begin{thebibliography}{99}
\medskip

\bibitem{vanKampen} N. G. van Kampen, {\sl Stochastic Processes in Physics and
Chemistry} (North-Holland, Amsterdam, 1981).

\bibitem{Haken} H. Haken, {\sl Synergetics} (Springer-Verlag, Berlin, 1978).

\bibitem{Nicolis} G. Nicolis and I. Prigogine, {\sl Self-Organization in
Non-Equilibrium Systems} (Wiley, New York, 1980).

\bibitem{Ligget} T. M. Liggett, {\sl Interacting Particle Systems}
(Springer-Verlag, New York, 1985).

\bibitem{reaction-reviews} K. Kang and S. Redner, Phys. Rev. A {\bf 32}, 435
(1985); V. Kuzovkov and E. Kotomin, Rep. Prog. Phys. {\bf 51}, 1479 (1988).

\bibitem{JStatPhys} J. Stat. Phys. Vol. {\bf 65}, nos. 5/6 (1991); this issue
contains the proceedings of a conference on {\sl Models of Non-Classical
Reaction Rates}, which was held at NIH (March 25--27, 1991) in honor of the
60th birthday of G. H. Weiss.

\bibitem{Laidler} K. J. Laidler, {\sl Chemical Kinetics}
(McGraw-Hill, New York, 1965).

\bibitem{Benson} S. W. Benson, {\sl The Foundations of Chemical Kinetics}
(McGraw-Hill, New York, 1960).

\bibitem{Coalescence} D. ben-Avraham, M. A. Burschka, and C. R. Doering, J.
Stat. Phys. {\bf 60}, 695 (1990).

\bibitem{Doering} C. R. Doering, Physica A {\bf 188}, 386 (1992).

\bibitem{waves} J. Riordan, C. R. Doering, and D. ben-Avraham, Phys. Rev. Lett.
{\bf 75}, 565 (1995).

\end{thebibliography}
\end{document}